\documentclass[12pt]{iopart}

\usepackage{cite}
\usepackage{blkarray}
\usepackage{xcolor}
\usepackage{multirow}
\usepackage{multicol}
\usepackage{bbold}
\usepackage{graphicx}
\expandafter\let\csname equation*\endcsname\relax
\expandafter\let\csname endequation*\endcsname\relax
\usepackage{amsmath}
\usepackage{iopams}  
\begin{document}

\title[Nonclassical features in higher-dimensional systems $\cdots$]{Nonclassical features in higher-dimensional systems through logical qudits}
  \author{Sooryansh Asthana$^{*}$ and V. Ravishankar$^{\dagger}$}
\address{Department of Physics, Indian Institute of Technology Delhi, New Delhi-110016, India.}
\ead{$^{*}$sooryansh.asthana@physics.iitd.ac.in, $^{\dagger}$vravi@physics.iitd.ac.in}
\vspace{10pt}
\begin{indented}
\item[]March 2023
\end{indented}

\begin{abstract}
 In a recent work [S. Asthana. New Journal of Physics 24.5 (2022): 053026], we have shown the interrelation of different nonclassical correlations in multiqubit systems   with quantum coherence in a single logical qubit. In this work,  we generalize it to higher-dimensional systems. For this, we take different choices of logical qudits and logical continuous-variable (cv) systems in terms of their constituent physical qudits and physical cv systems. Thereafter, we show reciprocity between conditions for coherence (in logical qudits and logical cv systems)  and conditions for nonlocality and entanglement (in their underlying constituent qudits and cv systems). 
   This shows that a single nonclassicality condition detects different types of nonclassicalities in different physical systems. Thereby, it reflects the interrelations of different nonclassical features of states belonging to Hilbert spaces of nonidentical dimensions. 
\end{abstract}
\vspace{2pc}
\noindent\submitto{\NJP}
\maketitle

\section{Introduction}
\label{Introduction}
Ever since the inception of quantum mechanics, there has been a burgeoning interest in the studies of quantum foundations \cite{einstein1935can, bell1964einstein, kochen1975problem, fine1982hidden, svetlichny1987quantum, werner1989quantum, mermin1990extreme, peres1996separability, ollivier2001quantum, wiseman2007steering, streltsov2017colloquium}. This is driven mainly by two reasons:  (i) the study of quantum foundations has unraveled many features that are at variance with their classical counterparts or they do not have any classical counterpart, (ii) resource-theoretic importance of different nonclassical features has been recognized in several quantum communication protocols, quantum computation, and quantum search algorithms \cite{ekert1991quantum, bennett1993teleporting, grover1997quantum, hillery2016coherence}, to name a few.

The prime examples of nonclassical correlations are quantum nonlocality (NL) and quantum entanglement \cite{einstein1935can, schrodinger1935discussion}. The resource-theoretic importance of nonclassical features has led to several  algebraic approaches for finding conditions for them. For example, many NL inequalities have been derived-- for multiqubit and multi-qudit systems (see, for example, \cite{brunner2014bell} and references therein). Their derivations are based on different local-hidden-variable (LHV) models \cite{svetlichny1987quantum, mermin1990extreme, seevinck2002bell, son2006test}, violations of classical probability rules \cite{collins2002bell}, and group-theoretic approaches (for example, stabiliser groups) \cite{guney2014bell, guney2015bell, gachechiladze2016extreme}. Similarly, the resource theory of entanglement has also been developed (see, for example, \cite{horodecki2009quantum} and references therein). Sufficiency conditions for entanglement, known as entanglement witnesses, have been derived through various approaches (see, for example, \cite{guhne2009entanglement}). In fact, the detection of different types of nonclassical correlations is ongoing research.

In parallel, the resource theory of quantum coherence has gained a lot of significance \cite{winter2016operational}.  The interrelation of quantum coherence with different nonclassical correlations, {\it viz.}, entanglement, nonlocality, and quantum discord has been studied from different  approaches \cite{Yu14quantum, Chitambar16, Sun2017quantum, qi2017measuring}.   In all these works, interrelations of nonclassical features have
been studied in the quantum systems belonging to Hilbert spaces of identical dimensions. This prompts us to ask the question: how are the nonclassical features of a mono-party lower-dimensional system related to that of a multi-party higher-dimensional system and vice-versa? This question is not without physical interest because, in fault-tolerant quantum computation, a single logical qubit (or qudit) is composed of  physical multiqubit (or multi-qudit) systems. In fact, different types of  logical qubits and qudits have been realized experimentally in the context of fault-tolerant quantum computation (such as superconducting qubits) and fault-tolerant quantum communication \cite{barends2014superconducting, chow2014implementing, muralidharan2014ultrafast, lee2020quantum}.

\begin{figure}[h]
 \centering  \includegraphics[width=10cm]{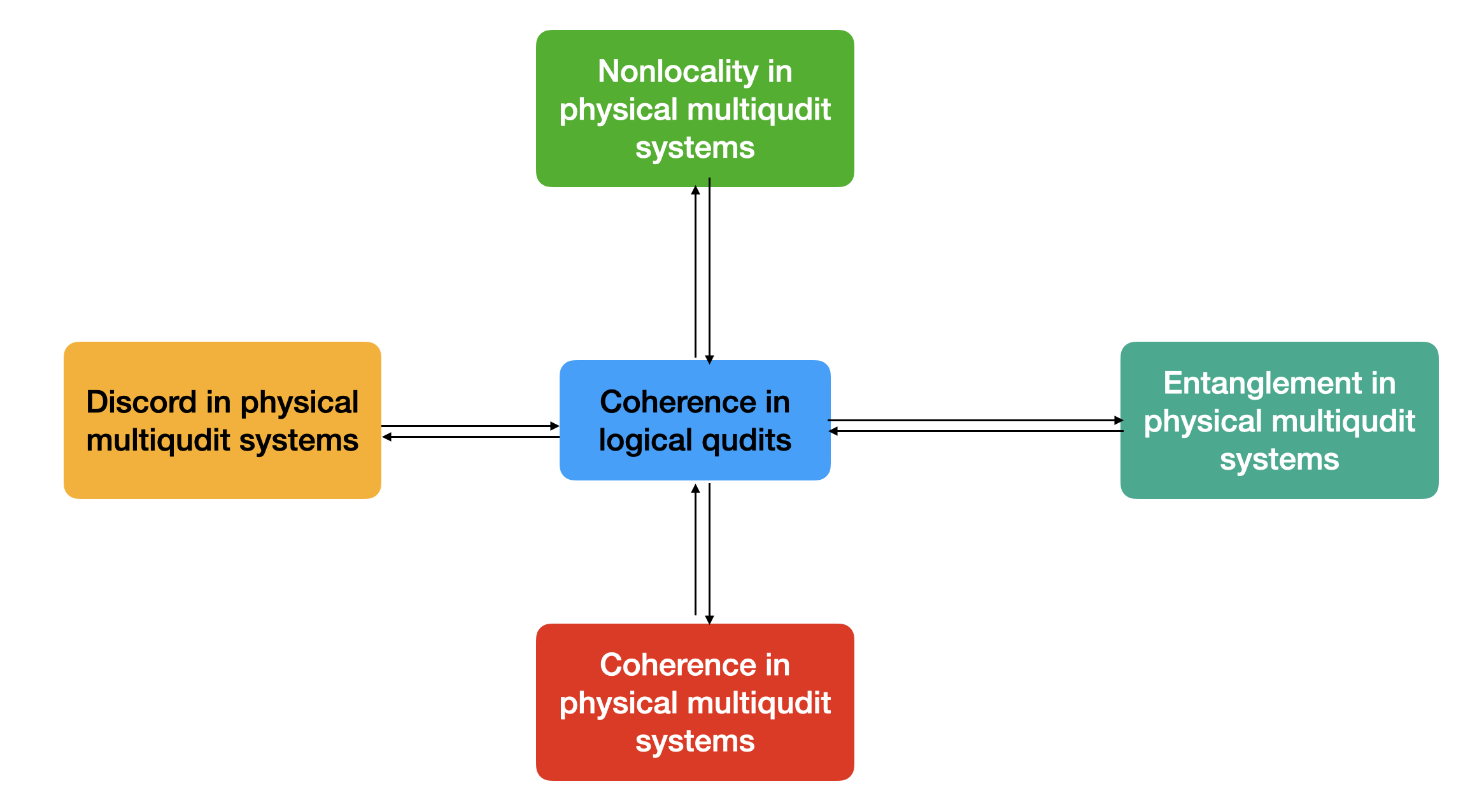}
    \caption{Coherence in logical qudits and nonclassical correlations in underlying physical multi-qudit systems. }
    \label{Idea}
\end{figure}

To answer this question, we have recently developed a methodology that employs homomorphic maps between stabilizer groups of a logical qubit and a physical multiqubit system as a tool.  The tools required to establish this mapping are fairly simple and not complicated. Logical bits have been used in classical error-correcting codes (for example, in parity codes) \cite{huffman2010fundamentals}. In contrast, quantum mechanics allows for many different types of logical qubits in terms of physical qubits, thanks to the principle of superposition \cite{calderbank1996good}. For this reason, coherence in a logical qudit is a manifestation of different types of nonclassical correlations in the underlying physical multi-qudit systems (depicted in figure (\ref{Idea}) and illustrated in section (\ref{Illustrations})). Hence, there should be  a mapping between the criteria for  quantum  coherence of logical qubits and those for nonclassical correlations (be it nonlocality or entanglement or quantum discord) of its physical constituent qubits and vice-versa. In  \cite{Asthana21interrelation}, we have shown that this mapping may be established by using the equivalence between logical qubits and different multiqubit physical systems.   Logical qudits and stabilizers have already been used extensively in the context of quantum error-correcting codes \cite{gottesman2002introduction}. To the best of our knowledge, they have not yet been employed to study the interrelations of different nonclassical features in different dimensions.

In this paper, we generalize  the methodology proposed in \cite{Asthana21interrelation} to multi-qudit systems and to continuous variable systems. The paper is structured as follows. Sections (\ref{Introduction}-\ref{Homomorphism}) consist of introduction and tools that we shall use in the rest of the paper. Sections (\ref{Only_procedure}-\ref{Continuous level system}) consist of applications, i.e., reciprocity among sufficiency conditions for different kinds of nonclassicalities and those for coherence. To elaborate,  we show that a single logical operator maps to many different operators acting on physical multi-qudit states (sections (\ref{Homomorphism}) and (\ref{Only_procedure})). Thereafter, we show reciprocity among sufficiency conditions for coherence in logical qudits and sufficiency conditions for nonclassical correlations in multi-qudit systems (sections (\ref{Framework}, \ref{Qutrit}, \ref{Coherence_to_GHZnonlocality}, \ref{Continuous level system})). Section (\ref{Conclusion_I}) concludes the paper.

%-------------------------------------------------

\section{Notation}
\label{Notation}
In this section, we set up the notations to be used throughout the paper.
\begin{enumerate}
    \item The subscript `$L$' is used for representing logical states or logical operators acting on the  Hilbert space of logical qudits. For example, the symbols $|\psi\rangle_L$ and $A_L$ represent a logical state and a logical operator respectively.
    \item The numeral subscript of an observable represents the party on which the observable acts. For example, the observables $A_1, \cdots, A_N$ act over the Hilbert spaces of the first, $\cdots$, the $N^{\rm th}$ qudit. 

    \item For qudit systems, the symbol $X_i$ represents the cyclic 
    translation operator by one unit acting on the $i^{\rm th}$ qudit. The symbol $Y_i$ represents the translation operator followed by a phase shift acting on the $i^{\rm th}$ qudit. The dimension in which these operators are defined will be clear from the context.
\end{enumerate}
\section{Different choices of logical qudits lead to different nonclassical correlations in multiqudit physical states}
\label{Illustrations}
We start with some simple observations to elucidate the interrelation between quantum coherence of logical quantum systems and underlying quantum correlations in its constituent physical subsystems:\\
\noindent {\it Observation 1: } Suppose that we  map multiqudit factorizable states (e.g., $|i\rangle^{\otimes N}$)  to  mono party logical qudit states (e.g., $|i\rangle_L$). In this case, the superposition of such multiqudit states maps to a coherent state in the  computational basis of logical states\footnote{At this juncture, we wish to stress that this clubbing is not just a mathematical artifice, its physical instances are manifest in atomic physics through the coupling of angular momenta, i.e., $L-S$ coupling and $jj$ coupling.}.\\
\noindent {\it Example:} The three-qudit entangled GHZ state, $|\psi_{\rm GHZ}\rangle \equiv \frac{1}{\sqrt{d}}\sum_{i=0}^{d-1}|iii\rangle$ maps to the maximally coherent logical state,  $|\psi_{\rm GHZ}\rangle_L \equiv \frac{1}{\sqrt{d}}\sum_{i=0}^{d-1}|i\rangle_L$, if $|iii\rangle \equiv |i\rangle_L$.

 This observation is valid for mixed states as well. It can be seen by invoking the convexity argument. Consider a mixed bipartite entangled state, $\rho = \sum_i p_i |\psi_i\rangle\langle \psi_i|$, in its eigenbasis. Since the state $\rho$ is  entangled, there exists at least one value of $i$ (say, $i=1$) for which the state $|\psi_1\rangle$ has a Schmidt rank $r> 1$, i.e., $|\psi_1\rangle \equiv \sum_{j=1}^r  \sqrt{\lambda_j}|jj\rangle$.  For the choice, $|jj\rangle \equiv |j\rangle_L$, $|\psi_1\rangle  \equiv \sum_{j=1}^r \sqrt{\lambda_j}|j\rangle_L$.  So, the state $\rho$ will exhibit coherence in the logical basis. This argument can be straightforwardly extended to multipartite states.  \\

\noindent {\it Observation 2:} The logical basis states involving superpositions of globally orthogonal but locally nonorthogonal physical states lead to different classes of entangled states. \\
{\it Example:} Let $|0\rangle_L\equiv |001\rangle, |1\rangle_L \equiv |010\rangle, |2\rangle_L\equiv |100\rangle$. The three logical states are globally orthogonal but locally nonorthogonal. Under this choice of logical qutrits,
\begin{align}
\frac{1}{\sqrt{3}}\Big(|0\rangle_L+|1\rangle_L+|2\rangle_L\Big) \equiv \frac{1}{\sqrt{3}}\Big(|001\rangle+|010\rangle+|100\rangle\Big).
\end{align}
On the other hand, if we choose $|0\rangle_L \equiv |000\rangle, |1\rangle_L\equiv |111\rangle, |2\rangle_L \equiv |222\rangle$, the same logical state gets mapped to a  three-qutrit state, 
  \begin{align}
     \frac{1}{\sqrt{3}}\Big(|0\rangle_L+|1\rangle_L+|2\rangle_L\Big) \equiv  \frac{1}{\sqrt{3}}\Big(|000\rangle+|111\rangle+|222\rangle\Big).
  \end{align}
  The difference between the three-qubit W-state and the three-qutrit GHZ state is that the former leads to an entangled state when one of its qubits is traced over.  The latter, however, leads to a fully separable state if any one of the qutrits is traced over.\\

\noindent {\it Observation 3:} Suppose that the logical basis states are built of entangled physical qudits. For this choice of logical states, a separable logical state may map to a physical multi qudit entangled state.\\
\noindent{\it Example:} Let $|0\rangle_L \equiv \frac{1}{\sqrt{2}}(|01\rangle-|10\rangle)$. For this choice of logical states, an incoherent separable state ({\it viz.}, $|00\rangle_L$) is a two-copy entangled two-qubit Bell state, $\Big(\frac{1}{\sqrt{2}}\big(|01\rangle-|10\rangle\big)\Big)^{\otimes 2}$.

\begin{table}[h!]
 \begin{center}
    \centering
\begin{tabular}{ | c| c| c |c|}
\hline
{\bf Logical state} & {\bf Choice of}  & {\bf Physical}  & {\bf Nonclassical} \\
&{\bf logical qudit} & {\bf multi qudit state}  &{\bf correlation}\\
\hline\hline
 $\frac{1}{\sqrt{d}}\sum_{i=0}^{d-1}|i\rangle_L$ & $|i\rangle_L \equiv |i\rangle^{\otimes N}$& $\frac{1}{\sqrt{d}}\sum_{i=0}^{d-1}|i\rangle^{\otimes N}$ & Genuine $N$-party \\
 & & & entanglement\\
 \hline
$\Big(\sum_{i}\sqrt{\lambda_i}|i\rangle_L\Big)\Big(\sum_j\sqrt{\mu_j}|j\rangle_L\Big),$&$|i\rangle_L\equiv|i\rangle^{\otimes N}$ and & $\Big(\sum_{i}\sqrt{\lambda_i}|i\rangle^{\otimes N}\Big)$ & {Biseparable state}\\ 
$0\leq \lambda_i, \mu_j \leq 1,$ &$|j\rangle_L\equiv |j\rangle^{\otimes M}$ &$\Big(\sum_j\sqrt{\mu_j}|j\rangle^{\otimes M}\Big)$ & \\
$\sum_i\lambda_i=\sum_j\mu_j=1.$ & & &\\
\hline
$ \frac{1}{\sqrt{3}}\big(|0\rangle_L+|1\rangle_L+|2\rangle_L\Big)$& $|i\rangle\equiv |iii\rangle$ & $\frac{1}{\sqrt{3}}\Big(\sum_{i=0}^2|iii\rangle\Big)$ & Three-qutrit \\
& & & GHZ entanglement\\
\hline
$ \frac{1}{\sqrt{3}}\big(|0\rangle_L+|1\rangle_L+|2\rangle_L\Big)$& $|0\rangle_L\equiv |001\rangle,$ & $\frac{1}{\sqrt{3}}\Big(|001\rangle+|010\rangle$ & Three-qubit \\
&  $|1\rangle_L\equiv |010\rangle,$  & $+|100\rangle\Big)$& W state\\
& $|2\rangle_L \equiv |100\rangle$ & &\\
\hline
$|00\rangle_L$ & $|0\rangle_L\equiv \frac{1}{\sqrt{2}}(|01\rangle-|10\rangle)$ &  $\Big(\frac{1}{\sqrt{2}}(|01\rangle-|10\rangle)\Big)^{\otimes 2}$ & {Biseparable}\\
\hline
 \end{tabular}
 \vspace{0.3cm}
    \end{center}
    \caption{Reciprocity between different nonclassical correlations in physical multi-qudit systems and coherence in logical qudits depending on different choices of logical qudits.}
    \label{Reciprocity_1}
\end{table}

These observations have been compactly shown in table (\ref{Reciprocity_1}).
Hence, in this language, all the sufficiency conditions for nonlocality and entanglement (in general, for any nonclassical correlation) in multi-qudit physical systems (and multimode cv states) should emerge from those for coherence in   logical qudits (resp., cv single-mode logical states) and vice-versa. 

 With this preface, we first present homomorphic maps between the stabilizer groups of physical multi-qudit states and that of a single logical qudit state. Thereafter, we lay down a procedure for obtaining sufficiency conditions for nonclassical correlations in physical multi-qudit systems from that for quantum coherence in a single logical qudit.

 %------------------------------
 \section{Homomorphic mapping between the stabilizer groups of a single qudit and a multi-qudit system}
 \label{Homomorphism}
 In this section, we present homomorphic mappings among the stabilizer groups of a single logical qudit and entangled physical multi-qudit systems. This homomorphic map will later be used as a tool to study reciprocity between sufficiency conditions for different nonclassical correlations and a sufficiency condition for coherence in logical qudits.
 \subsection{Homomorphism between stabilizer group of a single qudit state and that of a two-qudit state}
 
 We start with the state,
 \begin{align}
    |\psi\rangle_L = \frac{1}{\sqrt{d}}\sum_{i=0}^{d-1}|i\rangle_L.
    \label{Logical}
\end{align}
The stabiliser group of the state $|\psi\rangle_L$ is given by,
\begin{align}
  G_L:  \{\mathbb{1}, X_L, X_L^2, \cdots, X_L^{d-1}\};~~X_L =\sum_{k=0}^{d-1}|k+1\rangle_L{}_L\langle k|,
\end{align}
where the addition is modulo $d$. 
We next consider the state $|\psi\rangle_2 = \frac{1}{\sqrt{d}}\sum_{i=0}^{d-1}|ii\rangle$, whose stabiliser group is given by,
\begin{align}
   & G_2: \{H_2, (X_1X_2)^kH_2;~~1\leq k \leq (d-1)\};\nonumber\\
    & H_2\equiv \{\mathbb{1}, Z_1^xZ_2^y;~~ x+y =0~ {\rm mod}~d, 1\leq x, y\leq d-1\}, Z=\sum_{k=0}^{d-1}\omega^k|k\rangle\langle k|,
\end{align}
where $\omega$ is the $d^{\rm th}$ root of identity. 
It may be easily verified that $H_2$ is a normal subgroup of $G_2$.
The homomorphic map from $G_2$ to $G_L$ is given by,
\begin{align}
    H_2 \to \mathbb{1}_L,~~
    (X_1X_2)^kH_2 \to X^k_L ~~(1\leq k \leq d-1).
\end{align} 
\subsection{Homomorphism between stabilizer group of a single qudit state and that of a multi-qudit state}
The stabilizer group of the state,
\begin{align}
    |\psi\rangle_N =\dfrac{1}{\sqrt{d}}\sum_{i=0}^{d-1}|i\rangle^{\otimes N},
\end{align}
is given by,
\begin{align}
   & G_N: \{H_N, (X_1X_2\cdots X_N)^kH_N;~~1\leq k \leq (d-1)\},\nonumber\\
   & H_N\equiv \Big\{\mathbb{1}, Z_1^{x_1}Z_2^{x_2}\cdots Z_N^{x_N}; \sum_{i=1}^N x_i= 0~{\rm mod}~d;~0\leq x_i\leq d-1,~\nonumber\\
   &{\rm with~the~condition~that~any~one}~x_i~ {\rm cannot~ be~ nonzero~ with~ all~ other}~ x_i=0\Big\}.
   \label{coset}
\end{align}
As before, $H_N$ is normal in $G_N$ as may be seen by employing the relation $ZX=\omega XZ$. The homomorphic map from $G_N$ to $G_L$ is given by,
\begin{align}
    H_N &\to \mathbb{1}_L;~ (X_1X_2\cdots X_N)^kH_N \to X^k_L.
\end{align}

We shall employ these homomorphic maps  to obtain sufficiency conditions for nonclassical correlations in physical multi-qudit systems (resp., two-mode cv systems), given a sufficiency condition for coherence in a single logical qudit system (resp., a single mode cv system).
\section{Nonunique resolutions of a single logical operator for  a given choice of logical state}
\label{Only_procedure}
From the preceding section, it is clear that there exist different operators acting on physical multi-qudit systems which map to a single logical operator via a homomorphic map. Suppose that a logical qudit state, expressed in the computational basis $\{|i\rangle_L\}$, is a direct product of states of $N$ physical qudits, i.e., 
\begin{align}
|i\rangle_L \equiv |i\rangle^{\otimes N};~ 0\leq i \leq (d-1).    
\end{align} For this choice of logical qudit, which is of main interest to us in this paper, we  lay down a procedure to identify different operators acting on physical multi-qudit states as follows (the proof of the existence of such operators has been given in \ref{Proof}).

We start with an operator ${\cal A}_L$ which has the following equivalent forms,
 \begin{align}
 {\cal A}_L=\sum_{i,j=0}^{d-1} a_{ij}^{(L)}|i\rangle_L{}_L\langle j|=\sum_{i, j=0}^{d-1}a^{(L)}_{ij}\Big(|i\rangle\langle j|\Big)^{\otimes N}.  
 \end{align}
  The steps to identify distinct sets of operators acting over  physical qudits, that have the same overlap with the state $|\psi\rangle_L$ (given in equation (\ref{Logical})) are given as follows:
  \begin{enumerate}
    \item 
By construction, the operator ${\cal A}_L$ has support in a subspace of the direct product space ${\cal H}_1\otimes {\cal H}_2\otimes \cdots\otimes {\cal H}_N$. We map ${\cal A}_L$ to a sum of direct product of the operators $A_{1i}, A_{2i}, \cdots, A_{Ni}$,
\begin{align}
    {\cal A}_L \to \sum_i d_i \prod_{\alpha=1}^N{A}_{\alpha i},~~ d_i \in \mathbb{R},
\end{align}
so that the following condition is satisfied,
\begin{align}
    \langle\psi_L|{\cal A}_L |\psi_L\rangle= \Big\langle \psi_L\Big|\sum_i d_i \prod_{\alpha=1}^N{A}_{\alpha i}\Big|\psi_L\Big\rangle.
\end{align}
The operator $A_{\alpha i}$ acts on the logical qudit labeled $\alpha$. The resolution is not unique. Consider, for example, the following prescription:
\begin{align}
    \big({\cal A}_L\big)_{ij} \neq 0 \implies\prod_{\alpha=1}^N\big(A_{\alpha}\big)_{ij} =\big({\cal A}_L\big)_{ij},
\end{align}
     where $\prod_{\alpha=1}^N\big(A_{\alpha}\big)_{ij} $ represents element-by-element multiplication.  Following this, ${\cal A}_L \to A_1A_2\cdots A_N$.  Note that this prescription  allows for many choices of sets of operators $A_1, A_2, \cdots, A_N$ for the same logical operator ${\cal A}_L$. The action of the operator ${\cal A}_L$ on the state $|\psi_L\rangle$ is identical to that of the  tensor product of operators $A_1, A_2, \cdots, A_N$.

       \item Let $\Big\{d^{(k)}_i, A_{\alpha i}^{(k)}\Big\}$ represent the $k^{\rm th}$ resolution of ${\cal A}_L$. Then, the operator ${\cal A}_L$ can be variously expressed as $\sum_{k, i} w_k \prod_{\alpha=1}^N d_i^{(k)}A_{\alpha i}^{(k)}$, where the sole condition on $w_k$ is that they are normalized weights adding upto one, i.e., 
       \begin{align}
       {\cal A}_L \to \sum_{k, i} w_k \prod_{\alpha=1}^Nd_i^{(k)} A_{\alpha i}^{(k)}.   
       \end{align}
\end{enumerate}
\noindent{\it Example:} Consider a  logical qutrit, $|\phi\rangle_L\equiv \frac{1}{\sqrt{3}}(|0\rangle_L+|1\rangle_L+|2\rangle_L)$. It is an eigenstate of the operator $X_L \equiv |1\rangle_L{}_L\langle 0|+|2\rangle_L{}_L\langle 1|+|0\rangle_L{}_L\langle 2|$. Suppose that each logical qutrit is built of two qutrits (i.e., $|i\rangle_L \equiv |ii\rangle$). Let,
\begin{align}
    X &\equiv |1\rangle\langle 0|+|2\rangle\langle 1|+|0\rangle\langle 2|,~~
    Z \equiv |0\rangle\langle 0|+\omega|1\rangle\langle 1|+\omega^2|2\rangle\langle 2|,
\end{align}
where $\omega$ is the cube root of identity.  Employing $|i\rangle_L\equiv |ii\rangle$, the state $|\phi\rangle_L$ acquires the form, $\frac{1}{\sqrt{3}}(|00\rangle+|11\rangle+|22\rangle)$.
Making use of the point (i), the logical operator $X_L$ maps to the following operators,
\begin{align}
{X}_L  \to X_1X_2, (X_1Z_1)(X_2Z_2^2), (X_1Z_1^2)(X_2Z_2).
\end{align}

 In the next section, we  present the methodology to obtain sufficiency conditions for nonclassical correlations, given a  sufficiency condition for quantum coherence.
\section{Methodology}
\label{Framework}
 In this section, we  present the methodology to obtain sufficiency conditions for nonclassical correlations, given a sufficiency condition for quantum coherence.  

\label{Procedure}
 Suppose that a sufficiency condition for quantum coherence in the basis, $\{|i\rangle_L\}$ is given as
\begin{align}
{\cal C}_L: \sum_{a=0}^{d-1}c_a\Big\langle X_L^a\Big\rangle^{b_a} > c, 
\end{align}
 with the stipulation that it is maximally obeyed by the normalized state
$|\psi\rangle_L \equiv \frac{1}{\sqrt{d}}\sum_{i=0}^{d-1}|i\rangle_L$. $b_a$ are nonnegative integers and $c_a$ are real numbers. The value of $c$ is set in such a way that the ensuing inequality is violated by all the incoherent states in the computational basis of logical states.  
The task is to infer the underlying quantum correlation in the physical multiqudit states. That is to say, we examine what happens to the correlation in the physical multi-qudit systems as we vary the value of $c$. This would require   distinct sets of operators acting over  physical qudits that give rise to the witness ${\cal C}_L$. We use a two-pronged approach for obtaining sufficiency conditions for nonclassical correlations. That is to say, these operators can be identified in one of the following two ways:
\begin{enumerate}
\item  We identify the operators acting on multiqudit systems through homomorphism between the stabilizer groups. Suppose that $g^{(a)}$ represents an element belonging to the coset $(X_1\cdots X_N)^aH_N$ (given in equation (\ref{coset})). We use the homomorphic maps in the reverse direction for each of the operators $X_L^a$, i.e., 

\begin{align}
            X_L^a &\to  \sum_{g^{(a)}\in (X_1\cdots X_N)^aH_N}w^{(a)}_{g}g^{(a)},~~~~~~~~~~~
     \sum_{g^{(a)}\in (X_1\cdots X_N)^aH_N} w^{(a)}_{g} =1.
\end{align}
\item We identify the operators acting on multiqudit systems by using the procedure given in section (\ref{Only_procedure})  without using the homomorphic map. We may employ the resolutions of these operators,  i.e.,
     \begin{align}
{X}^a_L\to \sum_{k, i}w^{(a)}_k\prod_{\alpha=1}^Nd_{i}^{(k)}\Big(X_{\alpha i}^{(k)}\Big)^a,
      \end{align}
      where the subscript $\alpha$ labels the qudit over which the operator acts and the superscript $k$ labels the resolution of the operator $X_L$. $w_k^{(a)}$ are the weights and $d_i^{(k)}$ are real numbers.
\end{enumerate}
 The operator $\sum_{a=0}^{d-1}c_a{X}^a_L$ can be variously expressed as,
   \begin{align}
     & \sum_{a=0}^{d-1}c_a{X}^a_L \to   \sum_{a=0}^{d-1}c_a\sum_{g^{(a)} \in (X_1\cdots X_N)^aH} w_g^{(a)}g^{(a)},\\
    &~~~~~~~~~~~~~~~~~~~~~~~~~~~~~~~~~~~~~~~~~{\rm OR}~~~~~~~~~~~~~~~~~~~~~~~~~~~~~~~~~~\nonumber\\
  &   \sum_{a=0}^{d-1}c_aX_L^a\to \sum_{a=0}^{d-1}c_a\sum_{k, i}w^{(a)}_k\prod_{\alpha=1}^Nd_i^{(k)}\Big(X_{\alpha i}^{(k)}\Big)^a.
       \end{align}
 The choices of the weights $w_k^{(a)}$  or $w_g^{(a)}$ and the bound on the ensuing inequality depend on the notion of classicality under consideration. At this juncture, some comments are in order:
 \begin{enumerate}
     \item  Of all the resolutions, those resolutions are of particular interest for us in which the operators acting over the same qudit are noncommuting. These resolutions bring out nonclassical correlations in the underlying physical system.
     \item  We note that nonlocality is a nonclassical feature not restricted to quantum mechanics, whereas quantum coherence (in probability amplitudes) is a nonclassical feature restricted to quantum mechanics. A sufficiency condition for quantum coherence provides us with appropriate observables. The bound on  their combinations is set after ensuring that the combination is obeyed by all the LHV models. 
     \item We have taken but a simple choice of logical qudits $|i\rangle_L \equiv |i\rangle^{\otimes N}$. The procedure, however, is amenable to other choices of logical states discussed at the beginning of the section (\ref{Framework}). 
 \end{enumerate}

We now employ this methodology to identify  observables for constructions of entanglement inequalities and nonlocality inequalities for multiqudit as well as infinite-dimensional systems.

\section{Reciprocity between quantum correlation in a two-qutrit system and coherence for a single logical qutrit}
\label{Qutrit}
\subsection{Logical qutrit consisting of a pair of qutrit ($|i\rangle_L \equiv |ii\rangle$)}
 Let the basis states in the Hilbert space of logical qutrits be $|0\rangle_L, |1\rangle_L, |2\rangle_L$ and ${X}_L \equiv \sum_{i=0}^2|i+1\rangle_L{}_L\langle i|$, where the addition is modulo 3.
 We choose a sufficiency condition for coherence,
 \begin{align}
 |\langle {X}_L \rangle|> c, c\in [0, 1), 
 \label{Cond_coherence_qutrit}
 \end{align}
  with respect to the basis $\{|0\rangle_L, |1\rangle_L, |2\rangle_L\}$.
 The state,
 \begin{align}
 |\phi\rangle_L\equiv \frac{1}{\sqrt{3}}\Big(|0\rangle_L+|1\rangle_L+|2\rangle_L\Big), 
 \label{phiL}
 \end{align}
  maximally obeys this sufficiency condition, as $\langle \phi_L|{X}_L|\phi_L\rangle =1$.  We now move on to show how the sufficiency condition (\ref{Cond_coherence_qutrit}) in a logical system, gives rise to sufficiency conditions for quantum correlations in  two-qutrit systems. 

Let each logical qutrit be composed of a pair of qutrits, i.e.,  
\begin{align}
|0\rangle_L \equiv |00\rangle, |1\rangle_L\equiv |11\rangle, |2\rangle_L\equiv |22\rangle.    
\end{align}
 So, the state $|\phi\rangle_L$ assumes the form    $|\phi\rangle_L \equiv \dfrac{1}{\sqrt{3}}\Big(|00\rangle+|11\rangle+|22\rangle\Big)$.
 Following the procedure described in section (\ref{Procedure}), the logical operator, $X_L$, can be mapped to the following  physical two-qutrit operator,
\begin{align}
{X}_L\to & w_1X_1X_2+w_2
      (X_1Z_1)(X_2Z_2^2)+w_3
       (X_1Z_1^2)(X_2Z_2);~~\sum_{i=1}^3w_i=1, ~~0\leq w_i\leq 1.
       \label{Pair_ops}
\end{align}
 The operators $X_1$ and $X_2$ of the first and the second qutrit  have the same forms in the computational bases as that of ${X}_L$ in the basis $\{|0\rangle_L, |1\rangle_L, |2\rangle_L\}$.  The operator $Z$ is defined as: $Z\equiv {\rm diag}~(1, \omega, \omega^2)$, where $\omega$ is the cube root of identity. Obviously, $\omega^3 =1$  and  $[X_1, Z_1]\neq 0$, $[X_2, Z^2_2]\neq 0.$
\subsubsection{Entanglement in a two-qutrit system}
  If we choose $w_1=w_2=0.5$ and $w_3=0$ in  equation (\ref{Pair_ops}), the sufficiency condition (\ref{Cond_coherence_qutrit})  for coherence in logical qutrits yields the following sufficiency condition: 
\begin{align}
|\langle X_1X_2+X_1Z_1X_2Z_2^2\rangle| >2c,
\label{Ent_cond_sooryansh}
\end{align}
which is indeed the sufficiency condition for entanglement in two-qutrit systems if $ \frac{1}{2}\leq c <1$. In fact, as we decrease the value of  $c$ in the interval $\Big[\frac{1}{2}, 1\Big)$, the coherence witness for a logical qutrit (\ref{Cond_coherence_qutrit}) and hence, the condition for entanglement   in the underlying two-qutrit systems (\ref{Ent_cond_sooryansh}) becomes more encompassing. 

\subsubsection{Nonclassical correlation in two different bases in a two-qutrit system}
The sufficiency condition (\ref{Cond_coherence_qutrit}) also yields the following two sufficiency conditions (if $w_1=1$ and $w_2=1$),
\begin{align}
    |\langle X_1X_2\rangle|> c~~{\rm and}~~ |\langle X_1Z_1X_2Z_2^2\rangle|>c,
\end{align}
 for all nonzero values of $c$. If these two sufficiency conditions are simultaneously satisfied, they detect states having nonzero  correlations in the following two eigenbases of locally noncommuting operators:
\begin{enumerate}
    \item ${\cal B}_1$: common eigenbasis of $X_1$ and  $X_2$
    \item  ${\cal B}_2$:  common eigenbasis of  $X_1 Z_1$ and $X_2Z_2^2$.  
\end{enumerate}

 In this manner, depending upon the choices of weights and bound on the ensuing inequalities, the sufficiency condition for coherence (\ref{Cond_coherence_qutrit}) leads  to sufficiency conditions for different nonclassical correlations in the underlying two-qutrit system. 
%-------------------------------
\begin{table}[]
    \centering
    \begin{tabular}{|c|c|c|c|c|c|}
    \hline
      \multirow{2}{*}{ $\boldsymbol{w_1}$} & \multirow{2}{*}{ $\boldsymbol{w_2}$} &  \multirow{2}{*}{$\boldsymbol{w_3}$} &  \multirow{2}{*}{$\boldsymbol{c}$ }&  \multirow{2}{*}{{\bf Condition}} &  {\bf Nonclassical}\\
         & & & & & {\bf  correlation}\\
         \hline
         \multirow{2}{*}{$\frac{1}{2}$} & \multirow{2}{*}{$\frac{1}{2}$} & \multirow{2}{*}{0} 
         & \multirow{2}{*}{$\frac{1}{2} \leq c < 1$} & \multirow{2}{*}{$\vert\langle X_1X_2+(X_1Z_1)(X_2Z_2^2)\rangle\vert>2c$} &\multirow{2}{*}{ Entanglement}\\
         &&&&&\\
         \hline
      1 & 0 & 0 & $0 \leq c <1 $ &$\vert\langle X_1X_2\rangle\vert >c$  & Correlation in  \\
             0 & 1 & 0 & $0 \leq c <1 $ & $\vert\langle (X_1Z_1)(X_2Z_2^2)\rangle\vert >c$  &  bases ${\cal B}_1$ and ${\cal B}_2$\\
              \hline
    \end{tabular}
    \caption{Table showing the values of $w_1, w_2, w_3$, range of $c$, emergent sufficiency condition for nonclassical correlation and type of nonclassical correlation}
    \label{Table: correlation}
\end{table}
  The ranges of $c$, the values of $w_i$, the emergent sufficiency condition for nonclassical correlation and type of nonclassical correlation are shown in table (\ref{Table: correlation}).

\subsection{Logical qutrit consisting of a triplet of qutrit ($|i\rangle_L \equiv |iii\rangle$)}
Let each logical qutrit be composed of a triplet of physical qutrits, i.e.,  
\begin{align}
|0\rangle_L \equiv |000\rangle, |1\rangle_L\equiv |111\rangle, |2\rangle_L\equiv |222\rangle.    
\end{align}
 So, the state $|\phi\rangle_L$ (given in (\ref{phiL})) assumes the form    $|\phi\rangle_L \equiv \dfrac{1}{\sqrt{3}}\Big(|000\rangle+|111\rangle+|222\rangle\Big)$.
 Following the procedure described in section (\ref{Procedure}), the logical operator, $X_L$, can be mapped to the following  physical qutrit operator,
\begin{align}
{X}_L\to & w_1X_1X_2X_3+w_2
      (X_1Z_1)(X_2Z_2)(X_3Z_3);~0\leq w_1, w_2 \leq 1;~~w_1+w_2=1.
       \label{Pair_ops_3}
\end{align}
As before, the operators $X_1$ and $X_2$ of the first and the second qutrit  have the same forms in the computational bases as that of ${X}_L$ in the basis $\{|0\rangle_L, |1\rangle_L, |2\rangle_L\}$.  The operator $Z$ is defined as: $Z\equiv {\rm diag}~(1, \omega, \omega^2)$, where $\omega$ is the cube root of identity. Obviously, $\omega^3 =1$  and  $[X_i, Z_i]\neq 0, i \in \{1, 2, 3\}.$
\subsection{Entanglement in a three-qutrit system}
  If we choose the operators, $X_1X_2X_3, ~X_1Z_1X_2Z_2X_3Z_3$ (given in equation (\ref{Pair_ops_3})), the sufficiency condition (\ref{Cond_coherence_qutrit}) yields the following sufficiency condition: 
\begin{align}
|\langle X_1X_2X_3+X_1Z_1X_2Z_2X_3Z_3\rangle| >2c,
\end{align}
which is indeed the sufficiency condition for entanglement in three-qutrit systems if $ \frac{1}{2}\leq c <1$. In fact, as we decrease the value of  $c$ in the interval $\Big[\frac{1}{2}, 1\Big)$, the coherence witness for a logical qutrit and hence, entanglement witness for the underlying three-qutrit systems becomes more encompassing.

\subsection{Nonclassical correlation in two different bases in a three-qutrit system}
The sufficiency condition (\ref{Cond_coherence_qutrit}) also yields the following two sufficiency conditions,
\begin{align}
    |\langle X_1X_2X_3\rangle|> c~~{\rm and}~~ |\langle X_1Z_1X_2Z_2X_3Z_3\rangle|>c,
\end{align}
 for all nonzero values of $c$. If these two sufficiency conditions are simultaneously satisfied, they detect states having nonzero  correlations in the following two eigenbases of locally noncommuting operators:
\begin{enumerate}
    \item ${\cal B}_1$: common eigenbasis of $X_1$,  $X_2$ and $X_3$.
    \item  ${\cal B}_2$:  common eigenbasis of  $X_1 Z_1$, $X_2Z_2$ and $X_3Z_3$.  
\end{enumerate}
\begin{table}[]
    \centering
    \begin{tabular}{|c|c|c|c|c|}
    \hline
      \multirow{2}{*}{ $\boldsymbol{w_1}$} & \multirow{2}{*}{ $\boldsymbol{w_2}$} &  \multirow{2}{*}{$\boldsymbol{c}$ }&  \multirow{2}{*}{\bf  Condition} &  {\bf Nonclassical}\\
          & & & & {\bf  correlation}\\
         \hline
         \multirow{2}{*}{$\frac{1}{2}$} & \multirow{2}{*}{$\frac{1}{2}$}  
         & \multirow{2}{*}{$\frac{1}{2} \leq c < 1$} & \multirow{2}{*}{$\vert\langle X_1X_2X_3+(X_1Z_1)(X_2Z_2)(X_3Z_3)\rangle\vert>2c$} &\multirow{2}{*}{ Entanglement}\\
         &&&&\\
         \hline
      1 & 0  & $0 \leq c <1 $ &$\vert\langle X_1X_2X_3\rangle\vert >c$  & Correlation in  \\
             0 & 1  & $0 \leq c <1 $ & $\vert\langle (X_1Z_1)(X_2Z_2)(X_3Z_3)\rangle\vert >c$  &  bases ${\cal B}_1$ and ${\cal B}_2$\\
              \hline
    \end{tabular}
    \caption{Table showing the values of $w_1$ and $w_2$, range of $c$, emergent sufficiency condition for nonclassical correlation and type of nonclassical correlation}
    \label{Table: correlation_1}
\end{table}
 In this manner, depending upon the choices of weights and bound on the ensuing inequalities, the sufficiency condition for coherence (\ref{Cond_coherence_qutrit}) leads  to sufficiency conditions for different nonclassical correlations in the underlying three-qutrit system.  The ranges of $c$, the values of $w_i$, the emergent sufficiency condition for nonclassical correlation, and the type of nonclassical correlation are shown in table (\ref{Table: correlation_1}).

As another illustration, we now apply this procedure to show how observables employed in SLK inequality \cite{son2006test} can be straightforwardly identified by using sufficiency conditions for quantum coherence in a single logical system. 

%------------------------------------------------------------------------------------------------

\section{Reciprocity between coherence witness and conditions for GHZ nonlocality for qudits in even dimensions}
\label{Coherence_to_GHZnonlocality}
We start with a brief recapitulation of GHZ nonlocality in higher dimensions for pedagogic purposes. For details, see, for example, \cite{Lee06} and \ref{GHZ_NL_even}. The observables $X$ and $Y$ can be written in the computational basis set, $\{|n\rangle\}$, as,
\begin{align}
\label{X_1}
    X = \sum_{n=0}^{d-1}|n+1\rangle\langle n|,~~~~
    {Y} = \omega^{-1/2}\Big(\sum_{n=0}^{d-2}|n+1\rangle\langle n|-|0\rangle\langle d-1|\Big).
\end{align}

 The nonlocality inequality for a tripartite system is given by,
\begin{align}
    \Big\vert\Big\langle X_1X_2X_3+\omega X_1Y_2Y_3+\omega Y_1X_2Y_3+\omega Y_1Y_2Y_3\Big\rangle\Big\vert > 3.
    \label{Bell_Inequality_SLK_1}
\end{align}
It is maximally satisfied by the three-qudit GHZ state:
\begin{align}
\label{GHZ}
    |\psi_{\rm GHZ}\rangle = \dfrac{1}{\sqrt{d}}\sum_{n=0}^{d-1}|nnn\rangle.
\end{align}
%--------------------------------------
  The nonlocality inequality for tripartite $d$--diemsnional system, derived in \cite{son2006test}, is given as,
\begin{align}
    & \dfrac{1}{4}\sum_{n=1}^{d-1}\Big(\Big\langle (X_1X_2X_3)^n+(\omega X_1Y_2Y_3)^n+(\omega Y_1X_2Y_3)^n+(\omega Y_1Y_2Y_3)^n\Big\rangle\Big) + {\rm c.c.}> \dfrac{3d}{4}-1,~~ d~{\rm even}.
    \label{SLK__}
    \end{align}

This procedure admits a straightforward generalization to $N$ qudits and for arbitrary $d$.

In this section,  we show that the operators employed in GHZ nonlocality emerge as different resolutions of shift operators acting on a single logical  qudit system. We first start with a three-party system.  
\subsection{Tripartite system }
Let us assume that,
\begin{align}
 |0\rangle^{\otimes 3} \equiv |0\rangle_L, \cdots,  |d-1\rangle^{\otimes 3}\equiv |d-1\rangle_L.
\end{align}
 So, the tripartite generalized GHZ state can be written as a mono party logical state,
\begin{align}
\label{Logical_11}
    |\psi_{\rm GHZ}\rangle_L \equiv \dfrac{1}{\sqrt{d}}\sum_{n=0}^{d-1}|n\rangle_L.
\end{align} 
We start with a sufficiency condition for quantum coherence in a single logical system,
\begin{align}
\label{Coherence_Logical}
    \vert\langle X_L\rangle\vert> c, ~~~c\in [0, 1).
\end{align}

If we employ the equality    $|i\rangle_L \equiv |i i i\rangle,~i \in \{0, \cdots, d-1\}$, the operators $X_L$ may be reexpressed as,
\begin{align}
    X_L &\equiv \sum_{i=0}^{d-1}\Big(|i+1\rangle\langle i|\Big)^{\otimes 3}.
\end{align}
The sets of tensor products of local operators to which these operators map to are as follows (by following the procedure laid down in section (\ref{Procedure})):
\begin{align}
    X_L \rightarrow &w_1X_1X_2X_3+w_2 \omega X_1Y_2Y_2+w_3 \omega Y_1X_2Y_3+w_4 \omega Y_1Y_2X_3,\nonumber\\
    & \sum_{i=1}^4w_i=1,~0\leq w_1, w_2, w_3, w_4\leq 1.
    \label{S3}
\end{align}
where $\omega$ is the  $d^{\rm th}$ root of identity and the operators $X$ and $Y$ are defined in equations (\ref{X_1}) respectively. If we choose these operators, the sufficiency condition (\ref{Coherence_Logical}) yields the following sufficiency condition (corresponding to $w_1=w_2=w_3=w_4=\frac{1}{4}$):
\begin{align}
    \Big\vert\big\langle X_1X_2X_3+\omega X_1Y_2Y_3+\omega Y_1X_2Y_3+\omega Y_1Y_2X_3\big\rangle\Big\vert > 4c.
    \label{SLK_3}
    \end{align}

For $c= \frac{3}{4},$ it reduces to the sufficiency condition (\ref{Bell_Inequality_SLK_1}). The sufficiency condition (\ref{Coherence_Logical}) also yields the following four sufficiency conditions (corresponding to $w_1=1, w_2=1, w_3=1, w_4=1$ respectively),
\begin{align}
    |\langle X_1X_2X_3\rangle|> c,~~ |\langle \omega X_1Y_2Y_3\rangle|>c,~~|\langle \omega Y_1X_2Y_3\rangle|>c,~~|\langle \omega Y_1Y_2X_3\rangle|>c,
\end{align}
 for all nonzero values of $c$. If these four conditions are simultaneously satisfied, they detect states having nonzero  correlations in the following four eigenbases of locally noncommuting operators:
\begin{enumerate}
    \item ${\cal B}_1$: common eigenbasis of $X_1$,  $X_2$ and $X_3$.
    \item  ${\cal B}_2$:  common eigenbasis of  $X_1$, $Y_2$ and $Y_3$. 
    \item ${\cal B}_3$: common eigenbasis of  $Y_1$, $X_2$ and $Y_3$. 
    \item ${\cal B}_4$: common eigenbasis of  $Y_1$, $Y_2$ and $X_3$. 
\end{enumerate}
\begin{table}[]
    \centering
    \begin{tabular}{|c|c|c|c|c|c|c|}
    \hline
      \multirow{2}{*}{ $\boldsymbol{w_1}$} & \multirow{2}{*}{ $\boldsymbol{w_2}$} &  \multirow{2}{*}{$\boldsymbol{w_3}$}& \multirow{2}{*}{$\boldsymbol{w_4}$} &  \multirow{2}{*}{$\boldsymbol{c}$ }&  \multirow{2}{*}{Condition} &  Nonclassical\\
         & & & & & &  correlation\\
         \hline
         \multirow{2}{*}{$\frac{1}{4}$} & \multirow{2}{*}{$\frac{1}{4}$} &\multirow{2}{*}{$\frac{1}{4}$} & \multirow{2}{*}{$\frac{1}{4}$} 
         & \multirow{2}{*}{$\frac{3}{4}$} & \multirow{2}{*}{equation (\ref{SLK_3})} &\multirow{2}{*}{ Nonlocality}\\
         &&&&& &\\
         \hline
      1 & 0 & 0& 0 & $0 \leq c <1 $ &$\vert\langle X_1X_2X_3\rangle\vert >c$  & Correlation in  \\
             0 & 1 & 0  & 0 & $0 \leq c <1 $ & $\vert\langle \omega X_1Y_2Y_3\rangle\vert >c$  &  bases ${\cal B}_1$, ${\cal B}_2$, ${\cal B}_3$ and ${\cal B}_4$\\
                 0 & 0 & 1 & 0& $0 \leq c <1 $ &$\vert\langle\omega  Y_1X_2Y_3\rangle\vert >c$  &  \\
                      0 & 0 & 1 & 0& $0 \leq c <1 $ &$\vert\langle \omega Y_1Y_2X_3\rangle\vert >c$  &   \\
              \hline
    \end{tabular}
    \caption{Table showing the values of $w_1, w_2, w_3, w_4$, range of $c$, emergent condition for nonclassical correlation and type of nonclassical correlation}
    \label{Table: correlation_2}
\end{table}
 The ranges of $c$, the values of $w_i$, the emergent sufficiency condition for nonclassical correlation, and the type of nonclassical correlation are shown in table (\ref{Table: correlation_2}).
\subsection{$N$--party system}
If we consider another sufficiency condition for coherence in the logical qudits, 
    \begin{align}
    & \sum_{n=1}^{d-1}\langle X_L^n\rangle +{\rm c.c.}> c, c\in \big[0, 2(d-1)\big).\label{Coherence_condNparty}
\end{align}
This sufficiency condition gets maximally satisfied by the logical state  (\ref{Logical_11}). Since the state (\ref{Logical_11}) is a permutationally invariant state, it is invariant under the operators $X_L, X_L^2, \cdots, X_L^{d-1}$, where
\begin{align}
    X_L^i = \sum_{n=0}^{d-1}|n+i\rangle_L{}_L\langle n|,~ 1\leq i \leq (d-1).
\end{align}
\begin{table}[]
    \centering
    \begin{tabular}{|c|c|c|c|c|c|c|}
    \hline
      \multirow{2}{*}{ $\boldsymbol{w^{(i)}_1}$} & \multirow{2}{*}{ $\boldsymbol{w^{(i)}_2}$}& \multirow{2}{*}{ $\boldsymbol{w^{(i)}_3}$} &  \multirow{2}{*}{$\boldsymbol{w^{(i)}_4}$} &  \multirow{2}{*}{$\boldsymbol{c}$ }&  \multirow{2}{*}{\bf Condition} &  {\bf Nonclassical}\\
         & & & & & & {\bf  correlation}\\
         \hline
         \multirow{2}{*}{$\frac{1}{4}$} &   \multirow{2}{*}{$\frac{1}{4}$}  & 
           \multirow{2}{*}{$\frac{1}{4}$} & \multirow{2}{*}{$\frac{1}{4}$} 
         & \multirow{2}{*}{$\frac{3d}{4}-1$} & \multirow{2}{*}{equation (\ref{SLK__})} &\multirow{2}{*}{ Nonlocality}\\
         &&&&&&\\
         \hline
           1 & 0 & 0 & 0 &$0 \leq c <1 $ &$\vert\langle (X_1X_2X_3)^i\rangle\vert >c$  & Correlation in  \\
             0 & 1  & 0& 0 & $0 \leq c <1 $ & $\vert\langle (\omega X_1Y_2Y_3)^i\rangle\vert >c$  &  bases ${\cal B}^{(i)}_1$, ${\cal B}^{(i)}_2$, ${\cal B}^{(i)}_3$ and ${\cal B}^{(i)}_4$\\
                 0 & 0 & 1 & 0 & $0 \leq c <1 $ &$\vert\langle(\omega  Y_1X_2Y_3)^i\rangle\vert >c$  &  \\
                      0 & 0 &0 & 1 &  $0 \leq c <1 $ &$\vert\langle (\omega Y_1Y_2X_3\rangle)^i\vert >c$  &   \\
              \hline
    \end{tabular}
    \caption{Table showing the values of $w_1, w_2, w_3$, range of $c$, emergent sufficiency condition for nonclassical correlation and type of nonclassical correlation}
    \label{Table: correlation_3}
\end{table}
 By following the procedure laid down in section (\ref{Procedure}), the sets of tensor products of local operators to which these operators map are found to be:

\begin{align}
    X_L^i &\rightarrow w^{(i)}_1(X_1X_2X_3)^i+w^{(i)}_2(\omega X_1Y_2Y_2)^i+w^{(i)}_3(\omega Y_1X_2Y_3)^i+w^{(i)}_4(\omega Y_1Y_2X_3)^i;~~1\leq i \leq (d-1),\nonumber\\
    &~~~~~~~ 0\leq w_j^{(i)} \leq 1,~~ \sum_{j=1}^4w^{(i)}_j=1.
    \label{S3}
\end{align}

If we choose these operators, the sufficiency condition (\ref{Coherence_condNparty}) maps to the following sufficiency condition (corresponding to $w^{(i)}_1=w^{(i)}_2=w_3^{(i)}=w^{(i)}_4=\frac{1}{4};~~\forall i$), 
\begin{align}
    & \dfrac{1}{4}\sum_{n=1}^{d-1}\Big(\Big\langle (X_1X_2X_3)^n+(\omega X_1Y_2Y_3)^n+(\omega Y_1X_2Y_3)^n+(\omega Y_1Y_2Y_3)^n\Big\rangle\Big) + {\rm c.c.}> c,
    \end{align}
    which reduces to SLK nonlocality for $c=\frac{3d}{4}-1$ ($d$ even).  The sufficiency condition (\ref{Coherence_condNparty}) also yields the following sufficiency conditions (corresponding to $w^{(i)}_1=1, w^{(i)}_2=1, w^{(i)}_3=1, w^{(i)}_4=1$ respectively),
\begin{align}
    |\langle (X_1X_2X_3)^i\rangle|> c,~~ |\langle (\omega X_1Y_2Y_3)^i\rangle|>c,~~|\langle (\omega Y_1X_2Y_3)^i\rangle|>c,~~|\langle (\omega Y_1Y_2X_3)^i\rangle|>c,
\end{align}
 for all nonzero values of $c$. If these conditions are simultaneously satisfied, they detect states having nonzero  correlations in the following eigenbases of locally noncommuting operators:
\begin{enumerate}
    \item ${\cal B}^{(i)}_1$: common eigenbasis of $X^i_1$,  $X^i_2$ and $X^i_3$.
    \item  ${\cal B}^{(i)}_2$:  common eigenbasis of  $X^i_1$, $Y^i_2$ and $Y^i_3$. 
    \item ${\cal B}^{(i)}_3$: common eigenbasis of  $Y^i_1$, $X^i_2$ and $Y^i_3$. 
    \item ${\cal B}^{(i)}_4$: common eigenbasis of  $Y^i_1$, $Y^i_2$ and $X^i_3$. 
\end{enumerate}

In a similar manner, the coherence witness underlying CGLMP inequality \cite{collins2002bell} can be obtained by employing the form
 of CGLMP inequality given in \cite{son2006testa}. 
%----------------------------------------------------------

%-------------------------------------------------

\section{Generalisation to continuous variable systems}
\label{Continuous level system}
 We now turn our attention to continuous-variable (cv) systems. We show how a sufficiency condition for entanglement for a bipartite cv system gives rise to a sufficiency condition for coherence in a single cv system. 

 As an example of two-mode squeezed states, consider the two-mode squeezed vacuum. The corresponding squeezing operator is described by $S_g=e^{g(a_1a_2-a^{\dagger}_1a_2^{\dagger})}$, with squeezing parameter $g$. Its action on two-mode vacuum state results in \cite{PhysRevA.31.3093}, 
\begin{align}
    |\psi_{\rm tms}\rangle= (1-\tanh^2g)^{\frac{1}{2}}e^{\tanh g a_1^{\dagger}a_2^{\dagger}}|00\rangle.
\end{align}
 The mean photon number in both the modes is $N=2{\rm tr}(a_1^{\dagger}a_1|\psi_{\rm tms}\rangle\langle \psi_{\rm tms}|)=2\sinh^2g$. Of particular interest to us are the following operators, 
 \begin{align}
     O^{\pm}(\theta_1, \theta_2) \equiv X_1^{\theta_1}\pm X_2^{\theta_2} = \frac{1}{\sqrt{2}}(a_1e^{i\theta_1}+a_1^{\dagger}e^{-i\theta_1})\pm\frac{1}{\sqrt{2}}(a_2e^{i\theta_2}+a_2^{\dagger}e^{-i\theta_2}).
 \end{align}
 At the values $\theta=0, \frac{\pi}{2}$, the two observables, $X_i^0, X_i^{\pi/2}$ become canonically conjugate.
 The variance 
of the observable $O^{\pm}(\theta_1, \theta_2)$ is given by,
\begin{align}
    V(O^{\pm}(\theta_1, \theta_2))_{\psi_{\rm tms}}=\cosh 2g\pm\sinh 2g \cos(\theta_1+\theta_2).
\end{align}
 Duan-Simon criterion provides us with a sufficiency condition for  entanglement in cv systems \cite{PhysRevLett.84.2722}. 
 The criterion states that, if the state were separable, it would have obeyed,
\begin{align}
    V\Big(O^+(\theta_1, \theta_2)\Big)_{\rm sep}+V\Big(O^-(\theta'_1, \theta'_2)\Big)_{\rm sep}\geq 2,~~{\rm for}~~\theta_1-\theta'_1=\theta_2-\theta'_2=\frac{\pi}{2}.
    \label{Ent_conditions}
\end{align}
 On the other hand, the two-mode squeezed vacuum obeys the relation, 
\begin{align}
    V\Big(O^-(0,0)\Big)_{\psi_{\rm tms}}+V\Big(O^+\Big(\frac{\pi}{2}, \frac{\pi}{2}\Big)\Big)_{\psi_{\rm tms}}=2e^{-2g}.
\end{align}
which is distinctly less than 2, except at $g=0$.
 We now find the underlying sufficiency condition for coherence in a logical system. This may be done in two steps:
 \begin{enumerate}
     \item  Suppose that $\Pi$ represents the projection operator onto the subspace $\{|nn\rangle\}$, i.e., $\Pi = \sum_{n=0}^{\infty}|nn\rangle\langle nn|$.  We define the projected  operators 
    as follows: \begin{align}
O_p^{1\pm}\big(\phi, \phi\big) \equiv \Pi\Big({ X}_1^{\phi}\pm { X}_2^{\Phi}\Big)\Pi, ~~O_p^{2\pm}(\phi, \phi)\equiv \Pi({ X}_1^{\phi}\pm {X}_2^{\Phi})^2\Pi.
    \end{align}
     \item We next reexpress $|nn\rangle\equiv |n\rangle_L$. This yields the mode logical operators. Under this mapping, the two-mode squeezed vacuum state maps to the state $\frac{1}{\cosh r}\sum_{n=0}^{\infty}(\tanh r)^n|n\rangle_L$. So, the sufficiency condition for coherence underlying the entanglement condition (\ref{Ent_conditions}) is given by,
 \begin{align}
    &\big\langle O^{2-}_p(0,0)\big\rangle- \big\langle O^{1-}_p(0,0)\big\rangle^2
    + \big\langle O^{2+}_p\big({\pi/2}, {\pi/2})\big\rangle-\big\langle O^{1+}_p\big({\pi}/{2}, {\pi}/{2}\big)\big\rangle^2\geq 2. 
 \end{align}
 \end{enumerate}
  
 In a similar manner, the coherence witness underlying entanglement witnesses for an  $N$- mode state may be identified.

\section{Conclusion}
\label{Conclusion_I}
In summary, we have laid down a methodology to obtain a sufficiency condition for entanglement, nonlocality, and different kinds of nonclassical correlations, given a sufficiency condition for quantum coherence. As an application, we have shown how the sufficiency condition for generalized GHZ nonlocality emerges from a sufficiency condition for coherence if the three qudits are treated as a single logical qudit. We have also applied formalism to continuous-variable systems and shown the reciprocity between coherence in logical cv systems and entanglement in physical cv systems. This work shows how different nonclassical features are related to each other through logical qudits. Therefore, we believe that all the observables that are employed for higher-dimensional quantum-error correcting codes \cite{nadkarni2021quantum} may also be used for the detection of entanglement in the corresponding states.

In this work, we have studied  the interrelations of nonclassicality features in quantum states belonging to Hilbert spaces of different dimensions.  It is because the space of operators forms a vector space and superoperators can be defined as acting over this space of operators. That constitutes an interesting study that will be taken up elsewhere. 

Furthermore, the interrelations of different monogamy relations of different quantifiers of a resource, e.g., coherence with those of another resource, e..g, entanglement, nonlocality, etc. can also be studied.   In fact, what this work seems to suggest is that mathematically the same nonclassicality  conditions detect different types of nonclassicalities in different physical systems. Finally, if we employ the simplest decimal-to-binary mapping, the resulting mappings will naturally lead to another set of hierarchical relations of nonclassicality conditions in multiqubit systems and multiqudit systems. These conditions will be significant in resource theory of irreducible dimensions \cite{Cong2017witnessing, Kraft18}. 
\section*{Acknowledgement}
It is a pleasure to thank Rajni Bala for fruitful discussions, various insights, and for carefully going through the manuscript. Sooryansh thanks the Council for Scientific and Industrial Research (Grant No. -09/086 (1278)/2017-EMR-I) for
funding his research.
\appendix
 \section{There always exists at least two locally noncommuting operators corresponding to a single logical operator.}
 \label{Proof}
 Suppose that $|\psi_j\rangle$ represents a physical two-qudit state. We start with a basis of logical qudits,
    \begin{align}
{\cal B}_L \equiv \Big\{|i\rangle_L \equiv \sum_{j} c_{ij}|\psi_j\rangle; \sum_j|c_{ij}|^2 =1, \forall i\Big\} \in {\cal H}_L.        
    \end{align}
 The logical qudits have been chosen in such a manner that their coherent superposition, $|\psi\rangle_L\equiv \sum_i|i\rangle_L$, yields an entangled physical two-qudit system. 
The operator ${\cal A}_L \in {\cal H}_L$ has the following form,
\begin{align}
    {\cal A}_L \equiv \sum_{ij}a_{ij}|i\rangle_L{}_L\langle j|.
\end{align}

Since the state, $|\psi\rangle_L$ is an entangled physical two-qudit state and by construction, it has support over the full Hilbert space ${\cal H}_L$. So, dim ${\cal H}_L< d^2$ ($={\rm dim}~ H^d\otimes H^d$). For this reason, there exist more operators than one that have an identical action on the state $|\psi\rangle_L$ as that of $A_L$. This is explicitly shown below:

Consider, for example, the projection operator $\Pi$ onto the subspace orthogonal to ${\cal  H}_L$, ( ${\rm rank}~ \Pi> 1$). Suppose that the orthogonal subspace  is spanned by $\{|{\bf 0}_1\rangle, |{\bf 1}_1\rangle\}$ and $\{|{\bf 0}_2\rangle, |{\bf 1}_2\rangle\}$ (the subscripts label the first and the second party). The projection operator $\Pi$  has  the following resolution,
\begin{align}
    \Pi = &|{\bf 0}_1\rangle|{\bf 0}_2\rangle\langle {\bf 0}_1|\langle {\bf 0}_2|+|{\bf 1}_1\rangle|{\bf 1}_2\rangle\langle {\bf 1}_1|\langle {\bf 1}_2|
    \end{align}
Suppose that $U_1, U_2$ represent local $SU(2)$ transformations in the subspace spanned by $\{|{\bf 0}_1\rangle, |{\bf 1}_1\rangle\}$ and $\{|{\bf 0}_2\rangle,|{\bf 1}_2\rangle\}$ respectively. 
\begin{align}
     \Pi =& U_1\otimes U_2\Big(|{\bf 0}_1\rangle|{\bf 0}_2\rangle\langle {\bf 0}_1|\langle {\bf 0}_2|+|{\bf 1}_1\rangle|{\bf 1}_2\rangle\langle {\bf 1}_1|\langle {\bf 1}_2|\Big)U_1^{\dagger}\otimes U_2^{\dagger}\nonumber\\ \equiv& |{\bf 0}'_1\rangle|{\bf 0}'_2\rangle\langle {\bf 0}'_1|\langle {\bf 0}'_2|+|{\bf 1}'_1\rangle|{\bf 1}'_2\rangle\langle {\bf 1}'_1|\langle {\bf 1}'_2|.
\end{align} 
So, two non-commuting operators,
\begin{align}
O_1 &\equiv A_L +\lambda_1|{\bf 0}_1{\bf 0}_2\rangle\langle {\bf 0}_1{\bf 0}_2|+\lambda_2|{\bf 1}_1{\bf 1}_2\rangle\langle {\bf 1}_1{\bf 1}_2|;\nonumber\\
O_2 &\equiv A_L +\mu_1|{\bf 0}'_1{\bf 0}'_2\rangle\langle {\bf 0}'_1{\bf 0}'_2|+\mu_2|{\bf 1}'_1{\bf 1}'_2\rangle\langle{\bf 1}'_1{\bf 1}'_2|,~~ (\lambda_i \neq \mu_i), 
\end{align}
 will have identical actions on the state $|\psi\rangle_L$. A similar argument can be given for logical states built of multiqudit physical states.
 %------------------------------------------------------------------------
 \section{Brief recapitulation of GHZ Nonlocality in arbitrary even dimensions}
\label{GHZ_NL_even}
In this appendix, we present a derivation of the inequality (\ref{Bell_Inequality_SLK_1}). The generalised GHZ state $|\psi_{\rm GHZ}\rangle$ (given in equation (\ref{GHZ})) is the eigenstate of the observable $ X_1X_2X_3$ with eigenvalue $+1$,
\begin{align}
\label{LHV_1}
    X_1 X_2X_3|\psi_{\rm GHZ}\rangle =|\psi_{\rm GHZ}\rangle.
\end{align}

By using the symmetry operations ({\it viz.,} translational and permutational invariance) for the generalized GHZ state $|\psi_{\rm GHZ}\rangle$, other observables can be constructed. One such operator is given by $\omega X_1Y_2Y_3$ ($\omega$ is the $d^{\rm th}$ root of identity). 
The operator ${Y}$ is given in eqaution (\ref{X_1}).
In a similar manner, the other two  observables, $ \omega Y_1 X_2 Y_3$ and $ \omega Y_1 Y_2 X_3$ can be obtained. The obtained observables respectively have been shown to satisfy \cite{Lee06},
\begin{align}
    {X}_1 {Y}_2{Y}_3|\psi_{\rm GHZ}\rangle &=\omega^{-1}|\psi_{\rm GHZ}\rangle,~~
    {Y}_1{X}_2{Y}_3|\psi_{\rm GHZ}\rangle &=\omega^{-1}|\psi_{\rm GHZ}\rangle,~~
    \label{Nonlocality}
        {Y}_1{Y}_2{X}_3|\psi_{\rm GHZ}\rangle &=\omega^{-1}|\psi_{\rm GHZ}\rangle.
\end{align}

 In \cite{Lee06}, the existence of an underlying LHV model is assumed, and the forms $X_{\alpha} = \omega^{x_{\alpha}}$ and $Y_{\alpha}=\omega^{y_{\alpha}}$ for the outcomes of $X$ and $Y$ have been employed. $x_{\alpha}$ and $y_{\alpha}$ are integers. The constraint of the values assumed by the variables $X_{\alpha}$ and $Y_{\alpha}$ (outcomes of the corresponding observables) for each qudit $\alpha$ to be consistent with an underlying LHV model gets converted to the following constraints \cite{Lee06}:
\begin{align}
    x_1+y_2+y_3 &\equiv -1~{\rm mod}~d,~~~
    y_1+x_2+y_3 \equiv -1~{\rm mod}~d,\nonumber\\
    y_1+y_2+x_3 &\equiv -1~{\rm mod}~d.
    \label{Cond2}
\end{align}
Adding these equations yields the following condition,
\begin{align}
\label{LHV_condition}
    x_1+x_2+x_3 \equiv -2(y_1+y_2+y_3)-3~{\rm mod}~d.
\end{align}
Since the outcomes of $X_{\alpha}$ are  $\omega^{x_{\alpha}}$, equating the powers of $\omega$ both sides in equation (\ref{LHV_1}) under the assumption of an underlying LHV model leads to the following equation,
\begin{align}
\label{X__}
    x_1+x_2+x_3 \equiv 0~{\rm mod}~d.
\end{align} 
For an even integer $d$, the RHS of equation (\ref{LHV_condition}) is always an odd integer modulo $d$ for arbitrary $y_{\alpha}$. In other words, for even $d$, no integer can satisfy  the equation, $2y+3 \equiv 0$ mod $d$ where $y = y_1+y_2+y_3$. This contradicts the condition (\ref{X__}) emergent from eq. (\ref{LHV_1}). 
That is to say, equations (\ref{LHV_condition}) and (\ref{X__}) can not be solved simultaneously for any integer value of $y$.

This gives rise to  a  Hardy-type condition for nonlocality without inequality \cite{Hardy93} for an arbitrary even-dimensional tripartite system.

Since the conditions (\ref{LHV_1}) and (\ref{Nonlocality}) can not be obeyed by any LHV model simultaneously, they can be used to construct a Bell inequality. The inequality is given in equation (\ref{Bell_Inequality_SLK_1}). Since under an LHV model, only three conditions out of four (given in equations (\ref{LHV_1}) and (\ref{Nonlocality})) can be satisfied. So, the LHV bound is $3$. This can be straightaway generalized to $N$--party systems, with each subsystem being $d$--dimensional, where $N$ is odd and $d$ is an even integer. In fact, the $N$-partite SLK inequality for even dimensions has been given in \cite{son2006test}.

Equations similar to (\ref{LHV_1}, \ref{Nonlocality}) for higher powers of these observables have been shown in  the next section for $d=4$ for the purpose of illustration.
%---------------------------------------------------------------
\subsection{SLK nonlocality for higher powers of observables}
\label{Appendix}
In this section, we show how higher powers of observables $X_i, Y_i$ (given in equation (\ref{X_1})) also give rise to nonlocality conditions (for a detailed discussion, we refer the reader to \cite{lee2006greenberger}). The analysis in this appendix is restricted to $d=4$ only. We note that for the observables, $X_i^2, Y_i^2$, the following set of eigenvalue equations holds:
\begin{align}
  & (X_1X_2X_3)^2|\psi\rangle = |\psi\rangle;~
    (\omega X_1Y_2Y_3)^2|\psi\rangle = |\psi\rangle\nonumber\\
   & (\omega Y_1X_2Y_3)^2|\psi\rangle = |\psi\rangle;~~
    (\omega Y_1Y_2X_3)^2|\psi\rangle = |\psi\rangle,
        \label{NL_4_2}
\end{align}
where $|\psi\rangle$ is the three-ququart GHZ state, $\frac{1}{2}\sum_{i=0}^3|iii\rangle$. If we assume an LHV model and denote the outcomes of $X_i, Y_i$ by $\omega^{x_i}, \omega^{y_i}$ and equating the powers of $\omega$ both sides, the following set of equations results:
\begin{align}
    2(x_1+x_2+x_3) &\equiv 0~ {\rm mod}~4;~~
2(1+x_1+y_2+y_3) \equiv 0~{\rm mod}~4\nonumber\\
2(1+y_1+x_2+y_3) &\equiv 0~{\rm mod}~4;~~
2(1+y_1+y_2+x_3) \equiv 0~{\rm mod}~4.
\end{align}

Adding all three equations except the first one, the following equation results:
\begin{align}
    2\{3+(x_1+x_2+x_3)+2(y_1+y_2+y_3)\}\equiv 0~{\rm mod}~4
\end{align}
We incorporate the first equation, $2(x_1+x_2+x_3)\equiv 0$, in this equation to get
\begin{align}
    6+4(y_1+y_2+y_3) \equiv 0 ~{\rm mod}~4,
\end{align}
which cannot be satisfied for any integer values of $y_1, y_2, y_3$. 

Similarly, for $X_i^3, Y_i^3$, the following set of eigenvalue equations holds:
\begin{align}
   (X_1X_2X_3)^3|\psi\rangle &\equiv  |\psi\rangle;~~~
    (\omega X_1Y_2Y_3)^3|\psi\rangle \equiv  |\psi\rangle\nonumber\\
    (\omega Y_1X_2Y_3)^3|\psi\rangle &\equiv  |\psi\rangle;~~~
    \label{NL_4_3}
    (\omega Y_1Y_2X_3)^3|\psi\rangle \equiv  |\psi\rangle.  
\end{align}
As before, assuming an LHV model and denoting the outcomes of $X_i, Y_i$ by $\omega^{x_i}, \omega^{y_i}$ and equating the powers of $\omega$ both sides, the following set of equations results:
\begin{align}
    3(x_1+x_2+x_3) &\equiv 0~ {\rm mod}~4;~~~
3(1+x_1+y_2+y_3) \equiv 0~ {\rm mod}~4\nonumber\\
3(1+y_1+x_2+y_3) &\equiv 0~ {\rm mod}~4;~~~
3(1+y_1+y_2+x_3) \equiv 0~ {\rm mod}~4.
\end{align}

Adding all but the first equation, the following equation results:
\begin{align}
    3\{3+(x_1+x_2+x_3)+2(y_1+y_2+y_3)\}=0
\end{align}
We incorporate the first equation, $3(x_1+x_2+x_3)=0$, in this equation to get
\begin{align}
    9+6(y_1+y_2+y_3) \equiv 0 ~{\rm mod}~4,
\end{align}
which cannot be satisfied for any integer values of $y_1, y_2, y_3$.  Similarly, it can be extended to higher values of $d$ by a suitable choice of $Y_i$.

\section*{Bibliography}
%\bibliography{bibliography.bib}
%\bibliographystyle{plainnat}

\end{document}